\documentclass[aps,twocolumn]{revtex4}
\def\be{\begin{equation}}
\def\ee{\end{equation}}
\def\ba{\begin{eqnarray}}
\def\ea{\end{eqnarray}}

\usepackage{amsmath,bbm}
\usepackage{amssymb}
\usepackage{mathrsfs}
\usepackage{mathtools}
\usepackage{amsthm}
\usepackage{xcolor}
\usepackage{verbatim}
\usepackage[pdftex]{graphicx}
\usepackage{hyperref}
\hypersetup{
	colorlinks=true,
	linkcolor=blue,
	filecolor=magenta,      
	urlcolor=cyan,
}

\DeclareFontFamily{U}{rsfs}{}         
\DeclareFontShape{U}{rsfs}{m}{n}{<5> rsfs5 <6><7> rsfs7          %
  <8><9><10><10.95><12><14.4><17.28><20.74><24.88> rsfs10}{}     %
\DeclareMathAlphabet{\mathfs}{U}{rsfs}{m}{n}                     %
 \usepackage{braket}

\newcommand{\RN}[1]{%
  \textup{\uppercase\expandafter{\romannumeral#1}}%
}

\usepackage{braket}

\usepackage{xcolor}
\usepackage{xspace}
\usepackage{ifthen}

\newcommand{\showcomments}{true}

\newcommand{\mar}[1]%
{\ifthenelse{\equal{\showcomments}{true}}%
{{\color{orange}{\small \textbf{m:} #1}}}{\xspace}}%

\newcommand{\ale}[1]%
{\ifthenelse{\equal{\showcomments}{true}}%
{{\color{magenta}{\small \textbf{a:} #1}}}{\xspace}}%

\newcommand{\car}[1]%
{\ifthenelse{\equal{\showcomments}{true}}%
{{\color{blue}{\small \textbf{c:} #1}}}{\xspace}}%

\begin{document}
\title{Detecting  Gravitationally Interacting Dark Matter with Quantum Interference}

\author{Marios Christodoulou${}^e$, Alejandro Perez${}^{\star a}$, Carlo Rovelli${}^{abcd}$}

\affiliation{\vspace{.25cm}${}^a$ AMU Universit\'e, Universit\'e de Toulon, CNRS, CPT, F-13288 Marseille, EU}
\affiliation{${}^b$ Department of Philosophy, University of Western Ontario, London, ON N6A 3K7, Canada}
\affiliation{${}^c$ The Rotman Institute of Philosophy, 1151 Richmond St.~N London  N6A5B7, Canada}
\affiliation{${}^d$ Perimeter Institute, 31 Caroline Street N, Waterloo ON, N2L2Y5, Canada} 
\affiliation{${}^e$ Institute for Quantum Optics and Quantum Information, Boltzmanngasse 3, 1090 Vienna, Austria}
\date{\today}

 \begin{abstract} \noindent In spite of the large astronomical evidence for its effects, the nature of dark matter remains enigmatic. Particles that interact only, or almost only, gravitationally, in particular with masses around the Planck mass---the fundamental scale of quantum gravity---are intriguing candidates. Here we show that there is a theoretical possibility to directly detect such 
 particles using highly sensitive gravity-mediated quantum phase shifts. In particular, we illustrate a protocol utilizing Josephson junctions. \end{abstract}


\maketitle

{\em Introduction.} There is clear astrophysical evidence that the energy density of the universe includes a large component---denoted ``dark matter" (DM)---which is not described by the Standard Model (SM) of particle physics, and whose nature remains mysterious \cite{Freese:2008cz,DM}.
WIMPs (Weakly Interacting Massive Particles), suggested by supersymmetric extensions of the SM, have been a favorite candidate, but negative results of several searches and the failure to detect supersymmetry in particle accelerators \cite{ss} have squeezed them to less appealing theoretic corners. Searches for other candidates, such as  Axions, predicted by extensions of the strong sector of the SM, are currently underway \cite{Squid3}.  

An intriguing DM candidate is provided by particles that interact only or almost only gravitationally and have a mass of the order of the Planck mass ($m_P\sim 20\mu g$): these could immediately account for the astrophysical evidence, where DM is precisely revealed by its gravitational interaction.  The problem with this candidate is that---precisely because of the weakness of the 
 gravitational interaction---direct detection is expected to be hard \cite{Carney}. 
So, in a sense, a most natural DM candidate is the hardest to detect.  In this letter, we point out that detection could actually be within reach  employing  quantum interference techniques. 

\smallskip

{\em Why Planck-mass particles.}  The Planck scale is the fundamental scale in quantum gravity \cite{QG}. It is plausible to expect stable or quasi-stable objects at this scale as part of the spectrum. There are arguments indicating that 
quantum gravity could predict such particles \cite{PSDM}
and stabilize Planck-mass black hole remnants at the end of the evaporation \cite{CF}.
Hawking radiation theory predicts small black holes to radiate intensely, but the Planck scale is outside the domain of validity of Hawking's theory, which does not take quantum gravity phenomena into account \cite{Hawking}. 
Stable or semi-stable Planck mass objects could therefore be a consequence of  quantum gravity. This is a DM candidate that does not require modifications of the SM or modifications of general relativity above the Planck length scale. 
Furthermore, the strength of the interaction of such particles, combined with the assumption of a sufficiently hot big bang, leads to a density of these objects at decoupling whose order of magnitude is compatible with the present dark matter density \cite{Barrau:2019cuo, Amadei:2021aqd, Bengochea:2024msf}.  
\smallskip

{\em Quantum phases.}  Inspired by recent developments in the area of table-top experiments involving gravity and quantum phenomena, and the surrounding theoretical debate  (see \cite{Christodoulou:2022mkf, MC} and references therein) here we point out that direct detection may not be out of reach. Specifically, we make use of the fact that quantum phases can encode information about tiny momentum transfer, even if the corresponding  displacement is too small to be detected. 

 We first consider an idealised detector where the center of detector mass is set in a superposition of locations. We then discuss a more concrete protocol which employs the collective quantum state of electrons in a suitable  arrangement of Josephson junctions as the sensitive probe.  

 \smallskip

\begin{figure}[b]
	{\includegraphics[height=6cm]{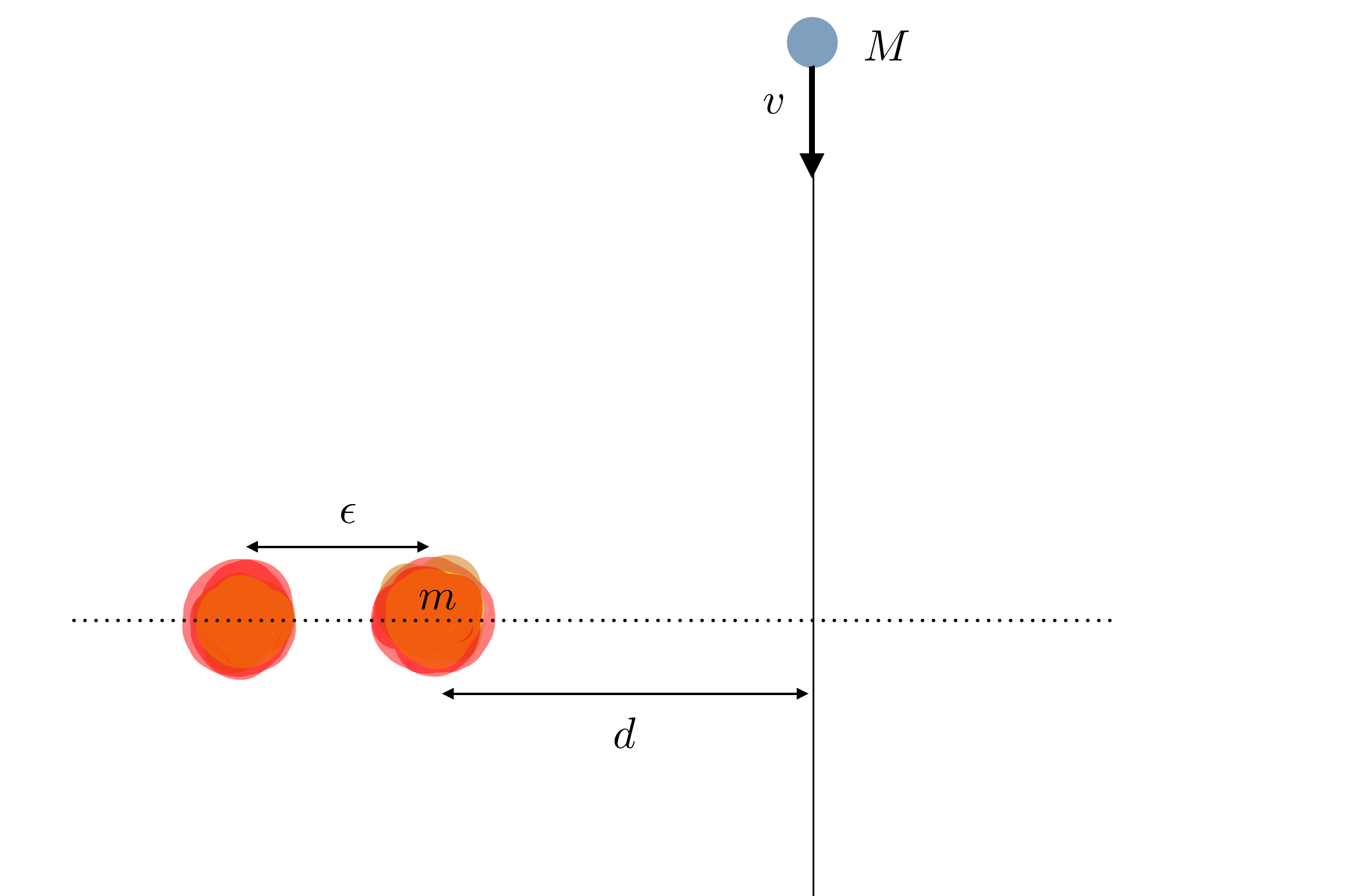}}
	\caption{A particle of mass $m$ in a superposition state with separation $\epsilon$. The DM particle passes by with velocity $v$.}.
	\label{AA-MB}
\end{figure}

{\em Idealized quantum protocol.}  Consider a quantum particle of mass $m$ (the ``detector'', or D particle) split into a superposition of two positions and then recombined. For concreteness, imagine it is a particle with spin $1/2$, prepared in the $|+\rangle_z$ eigenstate of the spin in the $z$ direction, and split according to the eigenstates $|\pm\rangle_y$ of the spin in the $y$ direction. Upon recombination, the particle will still be in the $|+\rangle_z$ state.  But say a (classical) particle with mass $M$ (the ``dark matter", or DM particle) flies rapidly next to one of the two positions, during the time the state was split.  The DM particle transfers different amounts of momentum to the two branches of the D particle, altering the relative phase. Upon recombination, the phase shift can give rise to a non-vanishing probability of measuring  $|-\rangle_z$.   Fig.\ref{AA-MB} illustrates the setting. 

The magnitude of the effect can be estimated as follows.  Take the D particle as the source of an external potential for the DM particle. 
As shown in the supplementary material (Eq.\eqref{displacement}), the displacement of the D particle during the passage of the DM particle is of the order of  
\be\label{four}
\Delta d\approx \frac{c^2}{v^2}\frac{M}{m_p} \ell_p,
\ee
where $d$ and $v$ are defined in Fig. 1 and $\ell_p$, $m_p$ and $c$ are the Planck length, the Planck mass, and the speed of light, respectively.  For $M\approx m_p$ and $v\approx 10^{-3} c$ (the mean velocity of DM particles in the galactic halo \cite{Freese:2012xd}) $\Delta d$ is of the order of $10^{6}\ell_p\sim 10^{-27 }cm$. Thus, detection using the classical response would be extremely hard (see \cite{Carney} for a detail analysis).

On the other hand, the relative \emph{quantum phase} between the two superimposed configurations can be estimated by evaluating the action difference between the two branches:
\be\label{ete}
\Delta S = \int dt \left(\frac{GmM}{\sqrt{d^2+(v t)^2}}-\frac{GmM}{\sqrt{(d+\epsilon)^2+(v t)^2}}\right),
\ee  
which only involves the difference of the integrated Newtonian potential in the two configurations of the D particle separated by the distance $\epsilon$. $G$ is the Newton constant. The integration of each term is logarithmically divergent, but the integration of the difference is finite. A direct evaluation gives 
\be\label{foury}
{\Delta S}= 2 \frac{G m M }{v} \log(1+\epsilon/d )\approx 2\frac{G m M}{v}\frac{\epsilon}{d}
\ee 
An improved calculation that takes into account the modification of the trajectory of the DM particle is given in the Supplementary Material. It changes the factor 2 in \eqref{foury} into a 3. 

The difference in the action gives a phase difference in the evolution of the two branches
\footnote{Notice that we have taken the gravitational field of the detector particle, and hence spacetime geometry, to be itself in quantum superposition, as pointed out in \cite{MC}.}
of the quantum state
\be\label{fase}
\Delta\phi=\frac{\Delta S}{\hbar}=3  \frac{{m}{M}}{m^2_{p}}\frac{c}{v}\frac{\epsilon}{d}.
\ee 
This result can also be understood as follows.   The difference of the action between the two branches is equal to the change of the Hamilton function for the motion of the DM particle in the field of the D particle. To first order, this is precisely encoded in the change in momentum by the general relation $\partial S/\partial x=-p$. Hence the above calculation can be seen as an evaluation of the difference in momentum transfer between the two branches \cite{Plenio}. This can be detectable even if the displacement in position is imperceptible.

A non-negligible phase shift, in fact, gives rise to a non-vanishing probability $P$ of measuring the recombined $D$ particle in the state $|-\rangle_z$, as 
\be\label{prob}
P
=\frac{1-\cos\Delta\phi}2.
\ee
If the dark matter particles have Planckian mass \cite{CF,Barrau:2019cuo,Amadei:2021aqd}, $M\sim m_{p}$,  then   
\be\label{def}
\Delta\phi \sim \frac{\epsilon }{d }\frac{c}{v}\frac{m}{m_{p}}. 
\ee 
 It is currently possible to  put a mass of the order  $m\approx 10^{-17} {\rm Kg}=2 \times10^{-8} m_p$ into quantum superposition \cite{Wien}.  The speed of cold DM particles in the galactic halo leads to an expected mean velocity on earth of $v\approx 10^{-3} c$ \cite{Freese:2012xd}. 
This gives $\Delta\phi \approx  10^{-5} \epsilon/ d$.
Due to the amplifying nature of the factor $c/v$ in eq. (\ref{def}), pushing technology to masses $m\sim 10^{-3} m_{p}$ is required, in order for the prefactor of $\epsilon/d$ to become order unity. This is far beyond current possibilities. Fortunately, a more realistic protocol can be designed using macroscopic collective quantum systems as for instance the one representing electrons in a superconductor.

\smallskip

\begin{figure}
	{\!\!\!\!\includegraphics[height=4.5cm]{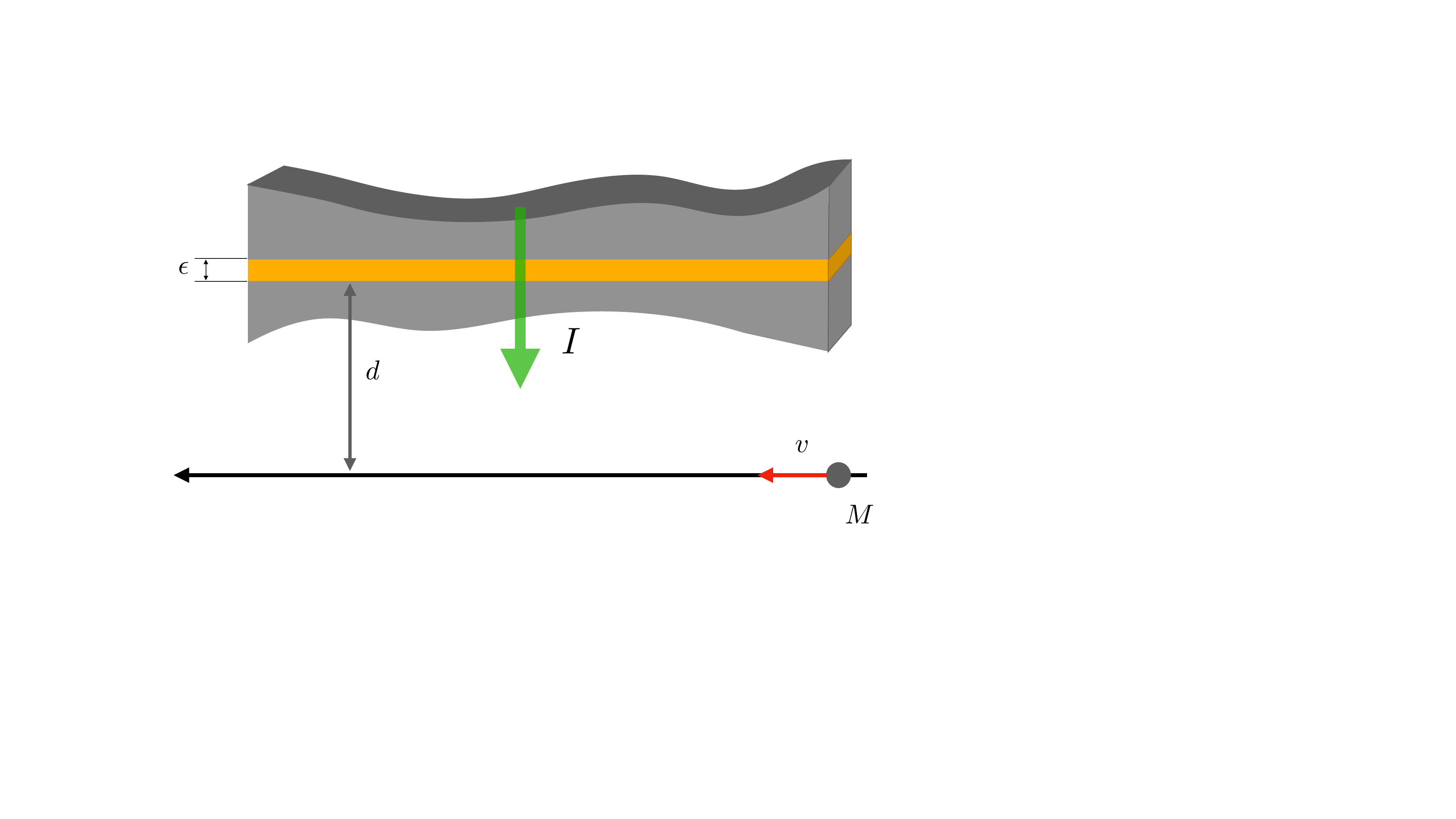}}
	\caption{An ensemble of aligned Josephson junctions working in parallel, interacting with the DM particle. This optimal can be achieved by suitably orienting the detector. }
	\label{josephson}
\end{figure}

{\em Josephson protocol.} The effect described above can be amplified when the phase shift (\ref{fase}) is induced in the wave function of a large number of particles in a coherent state. A device that allows to exploit this possibility is a superconducting Josephson junction (JJ)\footnote{Josephson junctions and SQUIDs have been suggested as super-sensitive gravitational detectors \cite{Squid1,Squid2} as well as for cold dark matter search \cite{Squid3}.}.  This realization of the detector has the advantage that the collective state of the electrons translates the probabilistic response of (\ref{prob}) into a directly measurable signal, circumventing the need of a statistical reconstruction of the phase.  
The standard theory of superconductors \cite{JJJ, bookJJ} yields two key equations  
\be
I=I_c \sin(\Delta\phi_e),\ \ \ \ \frac{\partial \Delta\phi_e}{\partial t}=\frac{\Delta\Phi}{\hbar}
\ee where $I$  is the electric current across the junction and $\Delta\Phi$ is the Cooper pairs potential energy difference between the two sides of the junction, $\epsilon$ is the insulator width of the JJ, and $I_c$ the critical current. 
The phase $\Delta\phi_e$ induced by the passage of a DM particle is given by (\ref{ete}) with $m=2m_e$. The spatial delocalization $\epsilon$ of the collective quantum state of the electrons is now given by the size of the insulating gap of the JJ, compare Figures 1 and 2. Using (\ref{fase}) and the previous numbers, we get $\Delta\phi_e\approx 10^{-19} \epsilon/d$ as the electron mass is $m_e\approx 10^{-22} m_p$.
For  small $\epsilon$ one has that $I_c\approx e\hbar n_s a /(m_e \epsilon)$ where $a$ is the area of the JJ and (at low temperatures) the density of superconducting electrons $n_s$ approaches the Fermi density $n_s\approx n_f=(3\pi^2)^{-1}(2m_e \varepsilon_f/\hbar^2)^{3/2}$ with $\varepsilon_f$ the Fermi energy of the material \cite{landau1980statistical}. Present technology allows for the integration of transistors close to the nanometer scale \cite{JJ} and it seems possible to produce JJ with $\sqrt{a} \sim 50 \, {\rm nm}$ in the future. Using this, and the value $\varepsilon_f\sim 7 eV$ (for copper) we estimate $I\sim (\epsilon/d) s^{-1}$ (electrons per second).  In an idealized aligned configuration of about $N\sim 10^8$ junctions connected in parallel along one meter (as in Fig \ref{josephson}) the DM particle would induce a current $I_{\rm out}=N I\approx 10^{7} \epsilon/d s^{-1}$ (electrons per second) with a single DM event ($I_{\rm out}\approx 10^{-11} (\epsilon/d) {\rm A}$).  

\smallskip

{\em Detector geometry.}
We illustrate for concreteness a possible macroscopic geometry for the detector. This will allow us to discuss 
the role of the parameters $d$ and $\epsilon$ in the estimates above, and the most relevant sources of external perturbations. Call a 1-dimensional line of JJs as described in the previous paragraph, and shown in Fig. \ref{josephson}, a ``detecting cell". The alignment of the detecting cell with the DM particle trajectory can be fixed by the astrophysical knowledge of the velocity field of DM in our galaxy. The actual detector can consist of a 3-dimensional array of  detecting cells (see Fig. \ref{detector}), arranged in a 2-dimensional  lattice, with separation $\ell$.  If appropriately oriented, a DM particle will maximally excite a few of the near neighbouring cells for which the impact parameter is $d \lesssim \ell$.  For the purposes of the present letter, we assume the apparatus can be designed so that $\ell\sim \epsilon$.  
Therefore, for estimates we can take $\epsilon/d\sim 1$.

\smallskip

{\em Noise.} Here are some estimates of the most obvious sources of noise, and some considerations on ways to deal with them.

A major source of noise is thermal. Interestingly, thermal noise in the output of a cell with $N$ JJs connected in parallel does not depend on $N$ and is given by $I_T\approx e kT/\hbar\approx 10^{-7} T/(1 {\rm K}) {\rm A}$ \footnote{Since the junctions are connected in parallel the current adds up. In the language of \cite{Carney},  this would correspond to $N$ independent measurements each seeing a tiny displacement. Here, we see only the cumulative additive effect.}. It is associated to the thermal fluctuations of the electron fluid in the superconductor. The signal to noise ratio is then $snr=I_{\rm out }=NI/I_T$.  For $N=10^8$, the condition $snr>1$ requires $T<10^{-1} {\rm mK}$. Much lower temperatures have been attained in the lab in small controlled environments \cite{Deppner:2021fks}, but achieving this at the space and time scales necessary for a realistic detector configuration is likely to be a key challenge.  WIMP-cryogenic detectors already operate at the ${\rm mK}$ regime \cite{Schumann:2019eaa},  but significant  development might be required to use this cooling technology for our setup as it is already exceedingly difficult for dilution
refrigeration to reach the single-digit mK regime.

A second source of noise is given by the macroscopic gravitational perturbations produced by nearby mass displacements (these can range from seismic modifications of the local gravitational field to the motion of massive bodies in the vicinity of the experimental setup). Phase differences due to nearly constant perturbations grow linearly in time at a rate given by the gradient of the gravitational potential (gravitational force) times $\epsilon$. This will produce a slow modulation of the current that can be distinguished from the fast DM particle signal with a time scale approximately given by their flying time $\Delta t=N\ell/v\sim 10^{-5} s$ {(in the setup used as illustration here)}.  In addition, macroscopic gravitational perturbations can be discerned from DM signals due to their global effect on the measuring cells. The smoking gun of a DM event being a local excitation of a few detecting cells along the track (Fig. \ref{detector}). Thus, it is reasonable to expect  macroscopic perturbations of the gravitational field to act on a very different time and space scale. This fact permits  the development of filtering techniques to extract the DM signal from this type of noise.

Flying-by charged particles constitute a standard noise source for regular particle detectors, such as those  used in searches for WIMPS. We expect standard techniques that deal with this issue to be applicable also to the present case. {Furthermore, note that if any charged particle does enter the detector (see next paragraph), it would excite a very large collection of cells due to the overwhelming strength of the electromagnetic interaction over the gravitational one. Therefore, we are in the setting where the weakness of gravity  compared to electromagnetism comes to our advantage as the local excitation of only a few cells should identify unambiguously a DM particle event. 

Electrically neutral particles can penetrate the detector and produce charged particles inside via ionisation. As in the previous case, experience with standard particle detectors suggest that this can also be under control in our setting.  Neutrons can be prevented from entering the devise using standard shielding techniques \cite{edelweiss}. Neutrinos cannot be shielded and can produce ionisation. But these  events are rare and do not seem to represent a problem. Using the data from \cite{XENON:2020kmp} one can estimate  less than $10^2$ ionisations per year in a one meter side detecting cube. Assuming that it takes one hour to reset the system, the probability that a DM particle crosses the detector and is missed due to such noise source is less than one event in $10^6$.

\smallskip

{\em Conclusion.} We have pointed out the  possibility that Planck mass particles making up dark matter that only interact gravitationaly could be detected using  quantum interference. We have illustrated an idealised setup where a mass is set in a superposition of locations, and a realistic setup using Josephson junctions. The central idea is to detect the relative quantum phase acquired by a macroscopic quantum system whose wavefunction is spatially delocalised. Concretely, we propose a large number of Josephson junctions  connected in parallel. Such arrangement allows for an amplification of the signal sufficient for a one-shot detection scheme. This seems necessary given the low rate of such events, as the flux of such DM particles on Earth is expected to be of the order of one particle per meter square per year \cite{Carney}. Covering a total area of several square meters with such detectors can give a significant rate. We have crudely estimated the most obvious sources of noise. The measurement may be within technological reach, but a more detailed feasibility study is needed. The challenge is significant, yet it is remarkable that quantum mechanics can amplify effects, which classically amount to hardly detectable Planck scale displacements (\ref{four}), to macroscopic observable levels. Rapidly evolving quantum  technologies combined with the growing interest in experiments testing the interface of gravity and quantum mechanics can be used to address  crucial questions in astrophysics, and possibly provide direct validation of certain implications of quantum gravity.

\begin{figure}
	{\includegraphics[height=6cm]{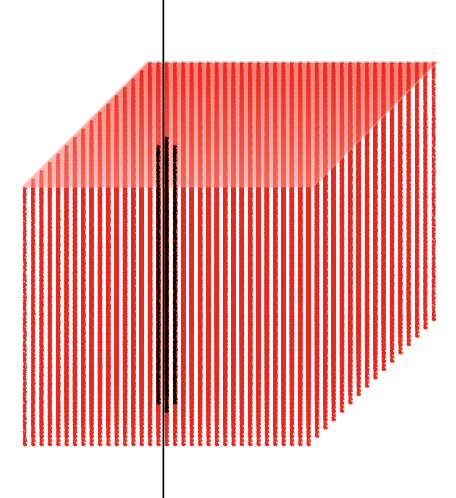}
 }
	\caption{Detecting cells 3d schematic configuration. A DM particle event excites local cells only (excited cells are represented as black lines).
 }
	\label{detector}
\end{figure}

\medskip

{\em Acknowledgements.} We thank G.
Higgins, T. Jonckheere, L. Raymond, J. Sofo, D. Sudarsky, A. Verga, and M. Zemlicka for fruitful discussions.
 We acknowledge support from the ID\# 61466 and ID\# 62312 grants from the John Templeton Foundation, as part of the ``Quantum Information Structure of Spacetime (QISS)'' project (\hyperlink{http://www.qiss.fr}{qiss.fr}).


\end{document}